\documentclass[pra,aps,twocolumn,showpacs,superscriptaddress]{revtex4}

\usepackage{amsmath}
\usepackage{amsfonts}
\usepackage{bm}
\usepackage{graphicx}

\begin{document}

\title{Dynamics of a vortex dipole across a magnetic phase boundary in a
spinor Bose--Einstein condensate}

\author{Tomoya Kaneda}
\affiliation{Department of Engineering Science, University of
Electro-Communications, Tokyo 182-8585, Japan}

\author{Hiroki Saito}
\affiliation{Department of Engineering Science, University of
Electro-Communications, Tokyo 182-8585, Japan}

\date{\today}

\begin{abstract}
Dynamics of a vortex dipole in a spin-1 Bose--Einstein condensate in which
magnetic phases are spatially distributed is investigated.
When a vortex dipole travels from the ferromagnetic phase to the polar
phase, or vice versa, it penetrates the phase boundary and transforms into
one of the various spin vortex dipoles, such as a leapfrogging
ferromagnetic-core vortex dipole and a half-quantum vortex dipole.
Topological connections of spin wave functions across the phase boundary
are discussed.
\end{abstract}


\maketitle

\section{Introduction}

A quantized vortex in a superfluid is a topological defect, and it
reflects the symmetry of the macroscopic wave function.
For example, a single-component Bose--Einstein condensate (BEC) is
described by a complex wave function $\psi \propto e^{i \phi}$ that has a
phase degree of freedom $\phi$, and its symmetry group is U(1).
As a result, a vortex is characterized by an integer winding number, known
as the Onsager--Feynman quantization~\cite{Onsager,Feynman}.
In a BEC of atoms with spin degrees of freedom, the ground-state manifold
has a more complicated topology, and a variety of spin vortices exist,
such as a half-quantum vortex~\cite{Leonhardt} and a polar-core
vortex~\cite{Sadler}.
Topological properties of spinor BECs have been studied by many
researchers~\cite{Choi,Ray,Ohmi,Ho,Khawaja,Savage,Ruo,Makela,Semenoff,Barnett,Kawaguchi,Michikazu,Pietila,Borgh,Kobayashi}.

The ground-state magnetic phase of a spin-1 BEC is the ferromagnetic or
polar phase depending on the interaction coefficients~\cite{Stenger}.
The symmetry group $G_f$ of the ferromagnetic phase is SO(3)~\cite{Ho},
i.e., there is a one-to-one correspondence between a state in the
ferromagnetic phase and an element in $G_f$.
The change in the spin state around a core of a spin vortex is therefore
equivalent to a closed loop in the $G_f$ manifold, which is classified by
the fundamental group $\pi_1(G_f) = \mathbb{Z}_2$.
For the polar phase, the symmetry group $G_p$ is ${\rm U}(1) \times S^2 /
\mathbb{Z}_2$, and its fundamental group is $\pi_1(G_p) =
\mathbb{Z}$~\cite{Makela}.

Let us consider a situation in which the magnetic phase containing a spin
vortex changes (spatially or temporally) from the ferromagnetic to polar
phases, or vice versa.
Since a change in the magnetic phase is accompanied by a change in its
symmetry group between $G_f$ and $G_p$, the topology of a spin vortex is
forced to change between elements in $\pi_1(G_f)$ and $\pi_1(G_p)$.
Such situations have been considered by several researchers.
In Ref.~\cite{Borgh}, connections of spin-vortex lines lying on both
sides of the two magnetic phases in a rotating spin-1 BEC were studied.
Tracing a vortex line from one side of the magnetic phase, it undergoes
various changes across the interface between the magnetic phases.
In Ref.~\cite{Kobayashi}, structures of spin-vortex cores were studied.
This problem is also regarded as the spin-vortex connection between
different phases, since the symmetry group of the spin state changes
radially from infinity to the vortex core.
In Ref.~\cite{Hoshi}, the sudden change in the interaction parameter from
polar to ferromagnetic was examined; a half-quantum vortex in the polar
phase was found to magnetize breaking the rotational symmetry.

In the present paper, we investigate the dynamics of a vortex dipole (a
vortex-antivortex pair) in a spin-1 BEC, where the spin-dependent
interaction parameter is spatially distributed, and the ferromagnetic and
polar regions are separated by a phase boundary.
We create a vortex dipole on one side of the magnetic regions, and it
moves toward the phase boundary.
When a vortex dipole has a sufficiently large velocity, it can penetrate
the phase boundary~\cite{Aioi}.
Passing through the phase boundary, the vortices experience a change in
the magnetic phase, and consequently, the topological properties of the
vortices are forced to change.
We will show that the system exhibits a rich variety of spin dynamics
that reflect the symmetry group of each magnetic phase.
When a singly-quantized vortex dipole moves from the ferromagnetic to the
polar phase, it transforms into a spin vortex dipole in which two
ferromagnetic cores exhibit a leapfrogging behavior, or it transforms into
a half-quantum vortex dipole.
When a singly-quantized vortex dipole moves from the polar to the
ferromagnetic phase, it transforms to a spin vortex dipole with
ferromagnetic cores surrounded by half-quantum vortices.

This paper is organized as follows.
Section~\ref{s:formulation} formulates the problem and provides a
numerical method.
Section~\ref{s:f_to_p} shows the dynamics of vortex dipoles traveling from
ferromagnetic to polar phases, and Sec.~\ref{s:p_to_f} shows those for
traveling from polar to ferromagnetic phases.
Section~\ref{s:conc} presents our conclusions from this study.

\section{Formulation of the problem}
\label{s:formulation}

We use mean-field theory to analyze a BEC of spin-1 atoms.
The macroscopic wave functions $\psi_m(\bm{r}, t)$ for magnetic sublevels
$m = 0, \pm1$ obey the Gross--Pitaevskii equation,
\begin{subequations} \label{GP}
\begin{eqnarray}
i\hbar \frac{\partial \psi_0}{\partial t} & = & -\frac{\hbar^2}{2M}
\nabla^2 \psi_0 + g_0 \rho \psi_0 + \frac{g_1}{\sqrt{2}} \left( F_+ \psi_1
+ F_- \psi_{-1} \right), \nonumber \\
& & \\
i\hbar \frac{\partial \psi_{\pm 1}}{\partial t} & = & -\frac{\hbar^2}{2M}
\nabla^2 \psi_{\pm 1} + g_0 \rho \psi_{\pm 1} 
\nonumber \\
& & + g_1 \left(
\frac{1}{\sqrt{2}} F_{\mp} \psi_0 \pm F_z \psi_{\pm 1} \right),
\end{eqnarray}
\end{subequations}
where $M$ is the atomic mass and $\rho = |\psi_1|^2 + |\psi_0|^2 +
|\psi_{-1}|^2$ is the atomic density.
The spin densities in Eq.~(\ref{GP}) are defined as $\bm{F} = \sum_{mm'}
\psi_m^* \bm{f}_{mm'} \psi_{m'}$, where $\bm{f}$ is the vector of spin-1
matrices, and $F_\pm = F_x \pm i F_y$.
The interaction coefficients in Eq.~(\ref{GP}) are given by $g_0 = 4 \pi
\hbar^2 (a_0 + 2 a_2) / (3M)$ and $g_1 = 4 \pi \hbar^2 (a_2 - a_0) /
(3M)$, where $a_0$ and $a_2$ are the $s$-wave scattering lengths with
colliding channels with total spin 0 and 2, respectively.
For simplicity, we focus on an infinite uniform system and Eq.~(\ref{GP})
contains no external potential terms.
We assume that the quadratic Zeeman effect is negligible.

The ground-state magnetic phase depends on the sign of the interaction
coefficient $g_1$.
For $g_1 < 0$, the spin-dependent interaction favors the ferromagnetic
ground-state.
Because of the U(1) gauge and spin rotation symmetries, the ground-state
manifold of the ferromagnetic state is expressed as~\cite{Ho}
\begin{equation} \label{ferro}
\zeta = e^{i\phi} R(\alpha, \beta, \gamma) \left( \begin{array}{c} 1 \\ 0
\\ 0 \end{array} \right) = e^{i(\phi - \gamma)} \left( \begin{array}{c}
e^{-i\alpha} \cos^2 \frac{\beta}{2} \\ \sqrt{2} \sin \frac{\beta}{2} \cos
\frac{\beta}{2} \\ e^{i\alpha} \sin^2 \frac{\beta}{2} \end{array} \right),
\end{equation}
where $e^{i\phi}$ and $R(\alpha, \beta, \gamma) = e^{-i f_z \alpha}
e^{-i f_y \beta} e^{-i f_z \gamma}$ are the U(1) and SO(3) rotations,
respectively.
Since $\phi$ can be absorbed into $\gamma$ in Eq.~(\ref{ferro}), the
symmetry group of the ferromagnetic state is $G_f =$ SO(3).
For $g_1 > 0$, the ground state is the polar phase given by
\begin{equation} \label{polar}
\zeta = e^{i\phi} R(\alpha, \beta, \gamma) \left( \begin{array}{c} 0 \\ 1
\\ 0 \end{array} \right) = e^{i\phi} \left( \begin{array}{c}
-\frac{1}{\sqrt{2}} e^{-i\alpha} \sin \beta \\ \cos\beta \\
\frac{1}{\sqrt{2}} e^{i\alpha} \sin \beta \end{array} \right).
\end{equation}
Since the state in Eq.~(\ref{polar}) is independent of $\gamma$ and
invariant with respect to $\alpha \rightarrow \alpha + \pi$, $\beta
\rightarrow \pi - \beta$, and $\phi \rightarrow \phi + \pi$, the symmetry
group of the polar state is $G_p = {\rm U}(1) \times S^2 / \mathbb{Z}_2$.

We consider a situation in which the interaction coefficient $g_1$ is
spatially distributed as
\begin{equation} \label{g2}
g_1(\bm{r}) = \left\{ \begin{array}{cc} g_f < 0 & (x < 0), \\
g_p > 0 & (x > 0), \end{array} \right.
\end{equation}
where $g_f$ and $g_p$ are negative and positive constants, respectively.
In the following, we will take $g_f = -0.01 g_0$ and $g_p = 0.01 g_0$.
Such a space-dependent interaction coefficient may be realized by, e.g.,
an optical Feshbach technique.
For the interaction coefficient $g_1$ in Eq.~(\ref{g2}), the ground state
is the ferromagnetic state for $x \rightarrow -\infty$ and the polar state
for $x \rightarrow \infty$.
Their phase boundary is located at $x \simeq 0$, where the two phases are
smoothly connected over the spin healing length.
When $\zeta|_{x \rightarrow -\infty} = (1, 0, 0)^T$, the spin state for $x
\rightarrow \infty$ must be $\zeta|_{x \rightarrow \infty} = (1, 0, 1)^T /
\sqrt{2}$ in the ground state, where the superscript $T$ stands for the
transpose.
Connection to other polar states, such as $\zeta|_{x \rightarrow \infty}
= (0, 1, 0)^T$, does not minimize the energy.

Spin vortices in a magnetic phase with symmetry group $G$ are
classified by its fundamental group $\pi_1(G)$.
For the ferromagnetic phase, the fundamental group of the symmetry group
is $\pi_1(G_f) = \mathbb{Z}_2$, and there are only two topological states:
no-vortex and vortex states.
An expression of the vortex state far from the core located at the origin
is given by $\zeta = (e^{i \theta}, 0, 0)^T$, where $\theta = {\rm arg}
(x + i y)$.
For the polar phase, $\pi_1(G_p) = \mathbb{Z}$, and corresponding vortex
states are written as $\zeta = (e^{i \theta}, 0, e^{i n \theta})^T /
\sqrt{2}$ with an integer $n$; this is a singly-quantized vortex for
$n = 1$ and a half-quantum vortex for $n = 0$.
The topological properties of spin vortices are thus different in the
regions $x < 0$ and $x > 0$ for the inhomogeneous interaction coefficient
$g_1$ in Eq.~(\ref{g2}).
In the following, we study the dynamics of a vortex dipole traveling
from $x < 0$ to $x > 0$, or vice versa.
We expect that the topology of the vortex dipole changes dynamically at
the interface between the ferromagnetic and polar phases.

We numerically solve Eq.~(\ref{GP}) using the pseudo-spectral
method~\cite{Recipes}.
The initial state is the ground state for the inhomogeneous interaction
coefficient $g_1$ in Eq.~(\ref{g2}).
The ground state is numerically obtained by the imaginary-time propagation
method, in which the $i$ on the left-hand side of Eq.~(\ref{GP}) is
replaced by $-1$.
In numerical simulations, a small numerical noise is added to the initial
state to break artificial symmetry.
A periodic boundary condition is imposed by the pseudo-spectral method.
The size of the system is taken to be sufficiently large, and the boundary
condition does not affect the dynamics of the vortices.

\section{Propagation of a vortex dipole from ferromagnetic to polar phase}
\label{s:f_to_p}

First we examine the dynamics of a vortex dipole traveling from the
ferromagnetic phase to the polar phase.
We prepare the ground state of Eq.~(\ref{g2}), where the atomic density
far from the phase boundary is $\rho_0$.
The spin state is $\zeta = (1, 0, 0)^T$ for $x \ll \xi_s$, and $\zeta =
(1, 0, 1)^T / \sqrt{2}$ for $x \gg \xi_s$, where $\xi_s = \hbar / (M g_1
\rho_0)^{1/2}$ is the spin healing length.
The spin healing length and the spin-wave velocity $v_s = (g_1 \rho_0 /
M)^{1/2}$ give the characteristic size and velocity of the spin dynamics,
which define the characteristic time scale $\tau = \xi_s / v_s$.
In the numerical simulations, a vortex dipole is created by imprinting the
phase on the wave functions as
\begin{equation} \label{phase}
\psi_m(z) \rightarrow \frac{(z - z_+) (z - z_-)^*}{|(z - z_+) (z - z_-)|}
\psi_m(z),
\end{equation}
on each numerical grid, where $z = x + i y$, and $z_\pm$ are the positions
of a vortex and an antivortex.
In the present initial state, we imprint the vortices at $z_\pm = -5 \xi_s
\pm i \Delta / 2$, where $\Delta$ is the distance between the vortex and
the antivortex.
The vortex dipole created on the ferromagnetic side ($x < 0$) then moves
in the $+x$ direction at a velocity $\simeq \hbar / (M \Delta)$.

\begin{figure}[tbp]
\includegraphics[width=8cm]{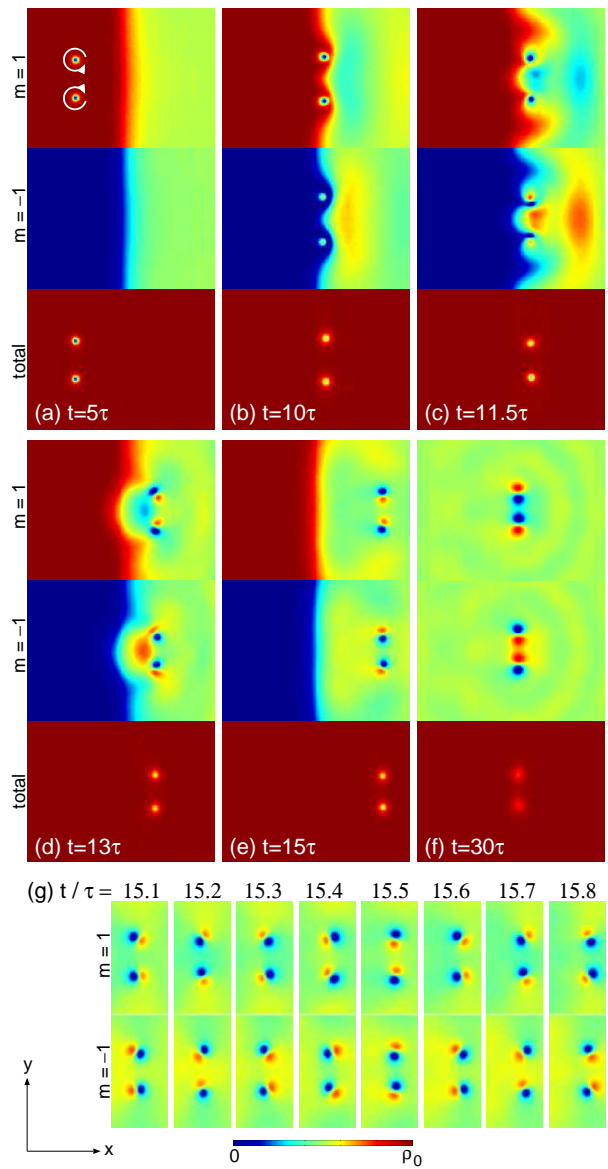}
\caption{
(Color online) Time evolution of a vortex dipole traveling from the
ferromagnetic phase to the polar phase, where the initial distance between
the vortices is $\Delta = 1.6\xi_s$.
(a)--(f) show $|\psi_1|^2$, $|\psi_{-1}|^2$, and $\sum_m |\psi_m|^2$, and
(g) shows $|\psi_1|^2$ and $|\psi_{-1}|^2$.
The density $|\psi_0|^2$ is always negligible.
The arrows in (a) indicate the directions in which the vortices circulate.
The windows of each panel are (a)--(e) $-4 \xi_s < x < 4 \xi_s$ and $-3
\xi_s < y < 3 \xi_s$, (f)$10 \xi_s < x < 18 \xi_s$ and $-3 \xi_s < y < 3
\xi_s$, and (g) $2 \xi_s < x < 4 \xi_s$ and $-2 \xi_s < y < 2 \xi_s$.
Note that the scale of (f) is the same as it is for (a)--(e).
See the Supplemental Material for a movie of the dynamics of
$|\psi_1|^2$.
}
\label{f:f_to_p1}
\end{figure}
Figure~\ref{f:f_to_p1} shows the dynamics of a vortex dipole for $\Delta
= 1.6 \xi_s$.
As the vortex dipole approaches the region of $x \simeq 0$, the phase
boundary is deformed, as shown in Fig.~\ref{f:f_to_p1}(b).
Passing through the phase boundary, the vortex dipole generates a
complicated spin texture, as shown in Figs.~\ref{f:f_to_p1}(c) and
\ref{f:f_to_p1}(d), and transforms into a spin vortex dipole traveling in
the polar phase, as shown in Fig.~\ref{f:f_to_p1}(e).
Despite the formation of the spin texture, the total density is almost
homogeneous except for that of the vortex cores.
Figure~\ref{f:f_to_p1}(g) shows in detail the vortex dynamics on the polar
side.
The cores of the vortex-antivortex pair in the $m = \pm 1$ components are
occupied by the $m = \mp 1$ components, and the vortex dipoles in both
components rotate around one another in a leapfrogging manner.
We note that this dynamics is similar to the well-known leapfrogging
behavior of vortices~\cite{Helmholtz}, but their mechanisms are quite
different; this will be discussed below.
As the vortex dipole travels in the polar phase, its cores gradually
expand, and the total density gradually increases, as shown in
Fig.~\ref{f:f_to_p1}(f).
This relaxation occurs as the leapfrogging behavior causes spin wave
radiation (see a movie in the Supplemental Material), dissipating the
energy of the vortex dipole.

\begin{figure}[tbp]
\includegraphics[width=8cm]{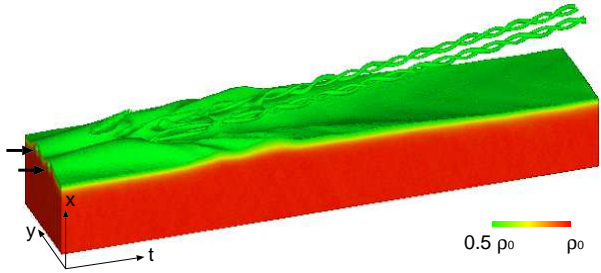}
\caption{
(Color online) Spatiotemporal image of the magnetization density
$|\bm{F}|$ during $10 < t / \tau < 16$.
The holes indicated by the arrows correspond to the vortex cores in
Fig.~\ref{f:f_to_p1}(b).
The parameters are the same as those in Fig.~\ref{f:f_to_p1}.
}
\label{f:mag}
\end{figure}
Figure~\ref{f:mag} shows the spatiotemporal image of the magnetization
density $|\bm{F}|$ for the dynamics in Fig.~\ref{f:f_to_p1}.
In the ferromagnetic region (red region in Fig.~\ref{f:mag}), the
magnetization is $|\bm{F}| / \rho_0 \simeq 1$ except for that of the
vortex cores (indicated by the arrows in Fig.~\ref{f:mag}).
After the penetration of the phase boundary, the vortex cores in the $m =
\pm 1$ components are occupied by the $m = \mp 1$ components, which
results in the magnetization $F_z / \rho_0 \simeq \mp 1$ at the vortex
cores.
The rotation of the vortex dipoles in Fig.~\ref{f:f_to_p1}(g) is thus
represented as double helices of magnetization in the spatiotemporal image
in Fig.~\ref{f:mag}.
The two helical structures in Fig.~\ref{f:mag} have opposite helicities.

The results shown in Figs.~\ref{f:f_to_p1} and \ref{f:mag} can be
understood as follows.
A singly-quantized vortex in the ferromagnetic phase is written as
\begin{equation} \label{ferrov}
\left( \begin{array}{c} e^{\pm i\theta} \\ 0 \\ 0 \end{array} \right)
f_1(r),
\end{equation}
where the origin of the polar coordinate is taken to be the center of the
vortex core, and $f_1(r) \geq 0$ is a radial function with $f_1(0) = 0$
and $f_1(r)|_{r \gg \xi} = \rho_0^{1/2}$ with $\xi = \hbar / (M g_0
\rho_0)^{1/2}$.
When the vortex in Eq.~(\ref{ferrov}) enters into the polar phase, it
transforms to
\begin{equation} \label{polarv}
\rightarrow \frac{e^{\pm i\theta}}{\sqrt{2}} \left( \begin{array}{c} 1 \\
0 \\ 1 \end{array} \right) f_{\rm p}(r) + \frac{1}{\sqrt{2}} \left(
\begin{array}{c} 1 \\ 0 \\ -1 \end{array} \right) f_{\rm p}^{\rm core}(r),
\end{equation}
where the radial functions $f_{\rm p}(r), f_{\rm p}^{\rm core}(r) \geq 0$
satisfy $f_{\rm p}(0) = 0$, $f_{\rm p}(r)|_{r \gg \xi_s} = \rho_0^{1/2}$,
and $f_{\rm p}^{\rm core}(r)|_{r \gg \xi_s} = 0$, in such a way that
$f_{\rm p}^{\rm core}(r)$ occupies the core of $f_{\rm p}(r)$.
The vortex states in Eqs.~(\ref{ferrov}) and (\ref{polarv}) can be
smoothly connected by an intermediate state,
\begin{eqnarray} \label{connect}
& & e^{\pm i\theta} \left( \begin{array}{c} \cos\chi \\
0 \\ \sin\chi \end{array} \right) {\cal F}[\chi; f_1(r) \rightarrow
f_{\rm p}(r)] \nonumber \\
& & + \left( \begin{array}{c} -\sin\chi \\ 0 \\ \cos\chi
\end{array} \right) {\cal F}[\chi; 0 \rightarrow f_{\rm p}^{\rm core}(r)],
\end{eqnarray}
where ${\cal F}$ smoothly connects the two functions as $\chi$ changes
from 0 (ferromagnetic) to $\pi / 4$ (polar).
The spin state of the core (the second terms in Eqs.~(\ref{polarv}) and
(\ref{connect})) is taken to be such that the total density is independent
of $\theta$.
In fact, the total density is almost isotropic around the vortices, as
shown in Figs.~\ref{f:f_to_p1}(a)--\ref{f:f_to_p1}(f).

We consider the time evolution of Eq.~(\ref{polarv}).
Since the first and second terms in Eq.~(\ref{polarv}) may have different
energies, the time evolution of Eq.~(\ref{polarv}) is written as
\begin{equation} \label{polarvt}
\frac{e^{\pm i\theta}}{\sqrt{2}} \left( \begin{array}{c} 1 \\
0 \\ 1 \end{array} \right) f_{\rm p}(r) + \frac{1}{\sqrt{2}} \left(
\begin{array}{c} 1 \\ 0 \\ -1 \end{array} \right) f_{\rm p}^{\rm core}(r)
e^{-i\delta\omega t},
\end{equation}
where $\hbar \delta\omega$ is the energy difference between the core and
the surrounding components.
The magnetization of Eq.~(\ref{polarvt}) is calculated to be
\begin{equation} \label{mag}
|\bm{F}| = |F_z| = 2 f_{\rm p}(r) f_{\rm p}^{\rm core}(r)
|\cos(\pm \theta + \delta \omega t)|,
\end{equation}
which has peaks at $r \simeq \xi_s$ and $\pm \theta + \delta\omega t = 0$
and $\pi$.
Thus, a vortex that enters into the polar phase has two regions of nonzero
magnetization around the core, and these rotate around one another at a
frequency $\delta\omega$, resulting in the magnetization helices shown in
Fig.~\ref{f:mag}.
From Figs.~\ref{f:f_to_p1}(g) and \ref{f:mag}, $\delta\omega\tau / (2\pi)$
is found to be $\simeq -2$.
This indicates that the energy of the core (the second term in
Eq.~(\ref{polarv})) is smaller than that of the surrounding component (the
first term in Eq.~(\ref{polarv})), since the core is not fully occupied,
as shown in the total densities in Fig.~\ref{f:f_to_p1}, and hence the
interaction energy is short of $g_0 \rho_0$ at the core.

The rotating vortices shown in Fig.~\ref{f:f_to_p1}(g) can also be
understood to be the result of the vortex-vortex interaction in an
effectively two-component BEC.
In a system with inhomogeneous density, a vortex with circulation
$\bm{\kappa}$ moves in the direction of $\nabla \rho \times \bm{\kappa}$,
i.e., in a direction perpendicular to the density
gradient~\cite{AioiX}.
Suppose that vortices in the $m = 1$ and $-1$ components are located at
the origin and in its vicinity, respectively.
The $m = \mp 1$ components occupy the vortex cores of $m = \pm 1$
components, and then the $m = -1$ density increases at the origin.
Therefore, when a clockwise (counterclockwise) vortex in the $m = -1$
component exists around the origin, it experiences a density gradient
toward the origin and moves around the origin clockwise
(counterclockwise).
Thus, clockwise (counterclockwise) vortices in the $m = 1$ and $-1$
components rotate around one another clockwise (counterclockwise), as
observed in Fig.~\ref{f:f_to_p1}(g).
A similar behavior can also be observed in a miscible two-component BEC
(data not shown).

\begin{figure}[tbp]
\includegraphics[width=8cm]{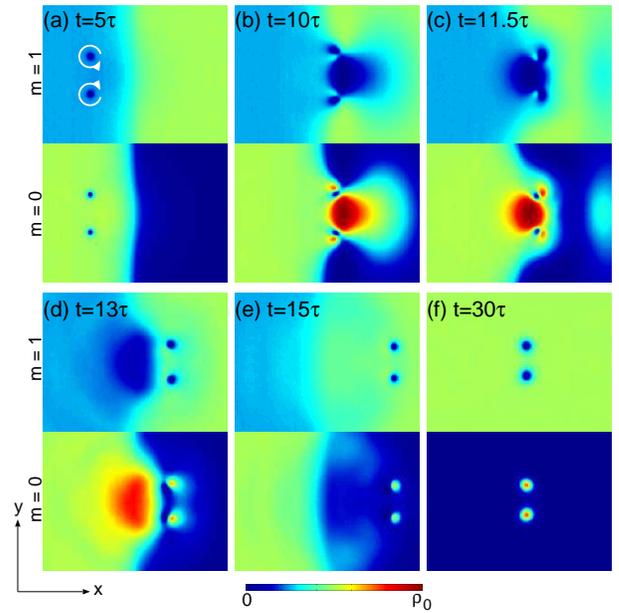}
\caption{
(Color online) The spin states in
Figs.~\ref{f:f_to_p1}(a)--\ref{f:f_to_p1}(f) are rotated by $\pi / 2$
about the $y$ axis.
The upper and lower panels show the density profiles
$|\psi_1|^2$ and $|\psi_0|^2$, respectively.
The density $|\psi_{-1}|^2$ is similar to $|\psi_1|^2$.
See the Supplemental Material for a movie of the dynamics of
$|\psi_1|^2$.
}
\label{f:f_to_p2}
\end{figure}
It is interesting to see the behavior in Fig.~\ref{f:f_to_p1} on a
different spin quantization axis.
Figure~\ref{f:f_to_p2} shows the same dynamics as in Fig.~\ref{f:f_to_p1},
where $e^{-i f_y \pi / 2}$ is applied to the spin state (rotation by $\pi
/ 2$ about the $y$ axis).
The ferromagnetic and polar states are rotated as $e^{-i f_y \pi / 2} (1,
0, 0)^T = (1/2, 1/\sqrt{2}, 1/2)^T$ and $e^{-i f_y \pi / 2} (1, 0, 1)^T /
\sqrt{2} = (1, 0, 1)^T / \sqrt{2}$.
After the vortex dipole penetrates through the phase boundary, it becomes
a vortex dipole in the $m = \pm 1$ components, and the cores are occupied
by the $m = 0$ component, as shown in Figs.~\ref{f:f_to_p1}(e) and
\ref{f:f_to_p1}(f).
Applying $e^{-i f_y \pi / 2}$ to Eq.~(\ref{polarvt}), we obtain
\begin{equation} \label{rotate}
\frac{e^{\pm i\theta}}{\sqrt{2}} \left( \begin{array}{c} 1 \\
0 \\ 1 \end{array} \right) f_{\rm p}(r) + \left( \begin{array}{c} 0 \\ 1
\\ 0 \end{array} \right) f_{\rm p}^{\rm core}(r) e^{-i\delta\omega t}.
\end{equation}
Unlike the behavior shown in Fig.~\ref{f:f_to_p1}(g), the rotation
dynamics of vortex dipoles is not observed with this spin quantization
axis.
The magnetization of Eq.~(\ref{rotate}) is of course the same as in
Eq.~(\ref{mag}).

\begin{figure}[tbp]
\includegraphics[width=8cm]{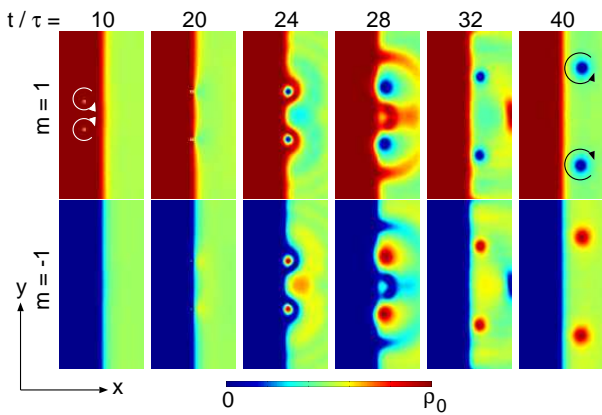}
\caption{
(Color online) Time evolution of a vortex dipole traveling from the
ferromagnetic phase to the polar phase, where the initial distance
between the vortices is $\Delta = 3.2\xi_s$.
The upper and lower panels show the density profiles $|\psi_1|^2$ and
$|\psi_{-1}|^2$, respectively.
The density $|\psi_0|^2$ is always negligible.
The arrows indicate the directions in which the vortices circulate.
The window of each panel is $-5 \xi < x < 5 \xi$ and $-10 \xi < y < 10
\xi$.
See the Supplemental Material for a movie of the dynamics of
$|\psi_1|^2$.
}
\label{f:hqv}
\end{figure}
Figure~\ref{f:hqv} shows the generation of a half-quantum vortex dipole,
where the distance in the initial vortex dipole $\Delta = 3.2 \xi_s$ is
larger than that in Figs.~\ref{f:f_to_p1}--\ref{f:f_to_p2}.
As they pass through the phase boundary, the size of the vortex cores with
the $m = 1$ components significantly expand and are occupied by the $m =
-1$ components, yielding the half-quantum vortex dipole in the polar
phase.
The transformation from a singly-quantized vortex in the ferromagnetic
phase to a half-quantum vortex in the polar phase is expressed as
\begin{eqnarray} \label{hqv1}
\left( \begin{array}{c} e^{\pm i\theta} \\ 0 \\ 0 \end{array} \right)
f_1(r) & \rightarrow &
\left( \begin{array}{c} e^{\pm i\theta} \cos\chi \\
0 \\ \sin\chi \end{array} \right) {\cal F}[\chi; f_1(r) \rightarrow
f_{\rm hqv}(r)] \nonumber \\
& & + \left( \begin{array}{c} 0 \\ 0 \\ 1 \end{array} \right)
{\cal F}[\chi; 0 \rightarrow f_{\rm hqv}^{\rm core}(r)]
\nonumber \\
& \rightarrow &
\frac{1}{\sqrt{2}} \left( \begin{array}{c} e^{\pm i\theta} \\
0 \\ 1 \end{array} \right) f_{\rm hqv}(r) + \left(
\begin{array}{c} 0 \\ 0 \\ 1 \end{array} \right)
f_{\rm hqv}^{\rm core}(r), 
\nonumber \\
\end{eqnarray}
where the radial functions $f_{\rm hqv}(r), f_{\rm hqv}^{\rm core}(r) \geq
0$ satisfy $f_{\rm hqv}(0) = 0$, $f_{\rm hqv}(r)|_{r \gg \xi_s} =
\rho_0^{1/2}$, and $f_{\rm hqv}^{\rm core}(r)|_{r \gg \xi_s} = 0$, i.e.,
$f_{\rm hqv}^{\rm core}(r)$ occupies the core of $f_{\rm hqv}(r)$.

Thus, when a vortex dipole created in the ferromagnetic phase enters into
the polar phase, it transforms into one of two kinds of spin-vortex
dipoles, as shown in Figs.~\ref{f:f_to_p1} and \ref{f:hqv}, depending on
the distance $\Delta$ in the initial vortex dipole.
The leapfrogging ferromagnetic-core vortex dipoles are generated for
$\Delta \lesssim 2 \xi_s$, and the half-quantum vortex dipole is generated
for $\Delta \gtrsim 2 \xi_s$.

From a topological point of view, in general, vortex states in the
ferromagnetic phase, classified by $\pi_1(G_f) = \mathbb{Z}_2$, and those
in the polar phase, classified by $\pi_1(G_p) = \mathbb{Z}$, can be
smoothly connected to each other~\cite{Kobayashi}.
In fact, spin states far from the core may be transformed from the
ferromagnetic phase to the polar phase as
\begin{equation} \label{general}
\left( \begin{array}{c} e^{\pm i\theta} \\ 0 \\ 0 \end{array} \right)
\rightarrow \left( \begin{array}{c} e^{\pm i\theta} \cos\chi \\ 0 \\
e^{i n \theta} \sin\chi \end{array} \right)
\rightarrow \frac{1}{\sqrt{2}} \left( \begin{array}{c} e^{\pm i\theta} \\
0 \\ e^{i n \theta} \end{array} \right),
\end{equation}
where $\chi$ changes from 0 to $\pi / 4$, and $n$ is an integer.
The dynamics shown in Figs.~\ref{f:f_to_p1} and \ref{f:hqv} correspond,
respectively, to $n = \pm 1$ and $n = 0$ in Eq.~(\ref{general}).
The transformation with other values of $n$ are not realized in the
present dynamics.

\section{Propagation of a vortex dipole from polar to ferromagnetic 
phase}
\label{s:p_to_f}

\begin{figure}[tbp]
\includegraphics[width=8cm]{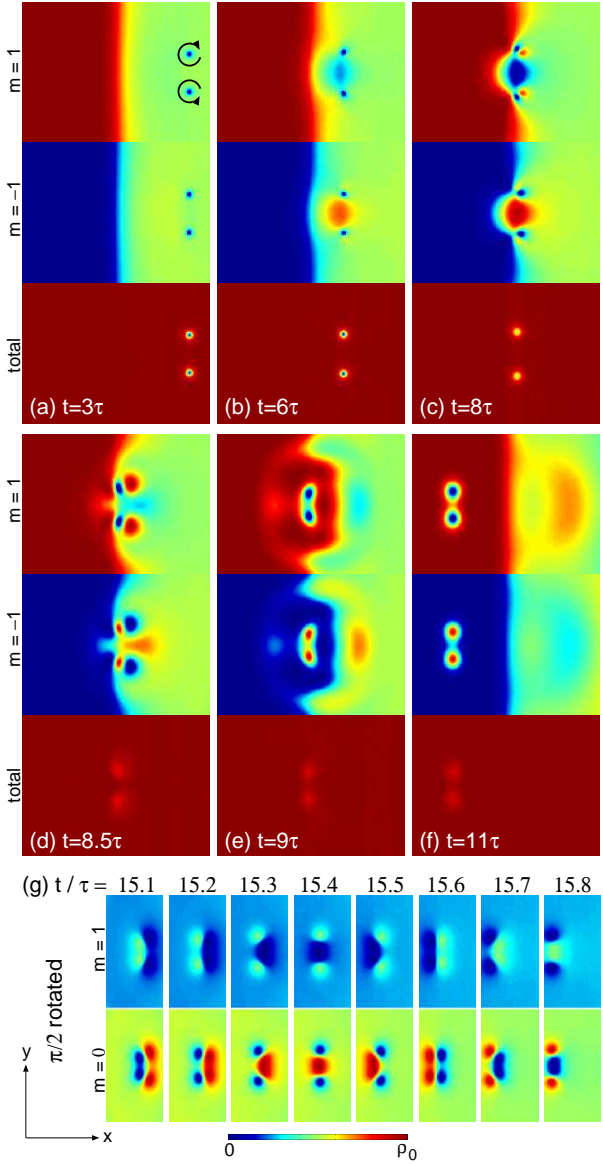}
\caption{
(Color online) Time evolution of a vortex dipole traveling from the
polar phase to the ferromagnetic phase, where the initial distance between
the vortices is $\Delta = 1.6\xi_s$.
(a)--(f) show $|\psi_1|^2$, $|\psi_{-1}|^2$, and $\sum_m |\psi_m|^2$,
where $|\psi_0|^2$ is always negligible.
The arrows in (a) indicate the directions in which the vortices circulate.
(g) Time evolution of the density profiles $|\psi_1|^2$ and $|\psi_0|^2$,
where the spin quantization axis is rotated by $\pi / 2$ about the $y$
axis.
The windows of each panel are (a)--(f) $-4 \xi_s < x < 4 \xi_s$ and $-3
\xi_s < y < 3 \xi_s$, and (g) $-4.5 \xi_s < x < -1.5 \xi_s$ and $-3 \xi_s
< y < 3 \xi_s$.
See the Supplemental Material for a movie of the dynamics of
$|\psi_{-1}|^2$ in (a)--(f) and that of $|\psi_1|^2$ in (g).
}
\label{f:p_to_f}
\end{figure}
We next consider cases in which a vortex dipole is created in the
polar phase ($x > 0$) and moves toward the ferromagnetic phase ($x < 0$).
In Fig.~\ref{f:p_to_f}, a vortex dipole is created at $z_\pm = 5 \xi_s \mp
i \Delta / 2$ with $\Delta = 1.6\xi_s$, which then moves in the $-x$
direction.
When the vortex dipole passes through the phase boundary, a complicated
spin texture is formed, as shown in Figs.~\ref{f:p_to_f}(c) and
\ref{f:p_to_f}(d).
After that, on the ferromagnetic side, a vortex dipole is transferred into
the $m = 1$ component, whose cores are occupied by the $m = -1$ component,
as shown in Fig.~\ref{f:p_to_f}(f).
Unlike the case in Fig.~\ref{f:f_to_p1}(f), the core size of the vortex
dipole traveling in the ferromagnetic phase is unchanged.
Figure~\ref{f:p_to_f}(g) shows the dynamics after passing through the
phase boundary, where the spin quantization axis is rotated by $\pi / 2$
about the $y$ axis ($e^{-i f_y \pi / 2}$ is applied).
In this quantization axis, the vortex dipole in Fig.~\ref{f:p_to_f}(f)
behaves as leapfrogging vortex dipoles in the $m = \pm 1$ and 0
components.

The vortex transformation from that in Fig.~\ref{f:f_to_p1}(a) to that in
Fig.~\ref{f:f_to_p1}(f) is expressed as
\begin{eqnarray} \label{pf}
\frac{e^{\pm i\theta}}{\sqrt{2}} \left( \begin{array}{c} 1 \\ 0 \\ 1
\end{array} \right) f_1(r) & \rightarrow &
e^{\pm i\theta} \left( \begin{array}{c} \cos\chi \\
0 \\ \sin\chi \end{array} \right) {\cal F}[\chi; f_1(r) \rightarrow
f_{\rm f}(r)] \nonumber \\
& & + \left( \begin{array}{c} -\sin\chi \\ 0 \\ \cos\chi
\end{array} \right) {\cal F}[\chi; 0 \rightarrow f_{\rm f}^{\rm core}(r)]
\nonumber \\
& \rightarrow &
e^{\pm i\theta} \left( \begin{array}{c} 1 \\ 0 \\ 0 \end{array} \right)
f_{\rm f}(r) + \left( \begin{array}{c} 0 \\ 0 \\ 1 \end{array} \right)
f_{\rm f}^{\rm core}(r), \nonumber \\
\end{eqnarray}
where $\chi$ changes from $\pi/4$ to 0.
The radial functions $f_{\rm f}(r), f_{\rm f}^{\rm core}(r) \geq 0$
satisfy $f_{\rm f}(0) = 0$, $f_{\rm f}(r)|_{r \gg \xi_s} = \rho_0^{1/2}$,
and $f_{\rm f}^{\rm core}(r)|_{r \gg \xi_s} = 0$.
In the final state in Eq.~(\ref{pf}), which corresponds to
Fig.~\ref{f:p_to_f}(f), the spin state has the form of $\propto
(e^{\pm i\theta}, 0, 1)^T$ for a radius $r \sim \xi_s$ satisfying
$f_{\rm f}(r) \simeq f_{\rm f}^{\rm core}(r)$, which is a half-quantum
vortex~\cite{Kobayashi}.
Thus, the vortex dipole in the ferromagnetic phase shown in
Fig.~\ref{f:p_to_f}(f) contains a half-quantum vortex dipole around the
ferromagnetic cores.

Applying $e^{-i f_y \pi / 2}$ to the final state in Eq.~(\ref{pf}) and
multiplying the second term by the factor $e^{-i \delta \omega t}$, we
obtain
\begin{eqnarray} \label{ferrovt}
& & e^{\pm i\theta} \left( \begin{array}{c} 1 / 2 \\
1 / \sqrt{2} \\ 1 / 2 \end{array} \right) f_{\rm f}(r) + \left(
\begin{array}{c} 1 / 2 \\ -1 / \sqrt{2} \\ 1 / 2 \end{array} \right)
f_{\rm f}^{\rm core}(r) e^{-i\delta\omega t}
\nonumber \\
& = & e^{-i\delta\omega t} \left( \begin{array}{c}
\left[
e^{i(\pm\theta + \delta\omega t)} f_{\rm f}(r) + f_{\rm f}^{\rm core}(r)
\right] / 2 \\ \left[
e^{i(\pm\theta + \delta\omega t)} f_{\rm f}(r) - f_{\rm f}^{\rm core}(r)
\right] / \sqrt{2} \\ \left[
e^{i(\pm\theta + \delta\omega t)} f_{\rm f}(r) + f_{\rm f}^{\rm core}(r)
\right] / 2 \end{array} \right).
\end{eqnarray}
Equation~(\ref{ferrovt}) has vortex cores at $r \sim \xi_s$ and
$\pm\theta + \delta\omega t = \pi$ in the $m = \pm 1$ component, and at
$\pm\theta + \delta\omega t = 0$ in the $m = 0$ component.
Thus, each component of Eq.~(\ref{ferrovt}) has a vortex core, and the
vortices in the $m = \pm 1$ and $m = 0$ components rotate around one
another with frequency $\delta\omega$, which explains the behavior in
Fig.~\ref{f:p_to_f}(g).
Since the positively (negatively) charged vortices rotate around one
another counterclockwise (clockwise) in Fig.~\ref{f:p_to_f}(g),
$\delta\omega$ is found to be negative, in a manner similar to the case of
Figs.~\ref{f:f_to_p1} and \ref{f:mag}.

\begin{figure}[tbp]
\includegraphics[width=8cm]{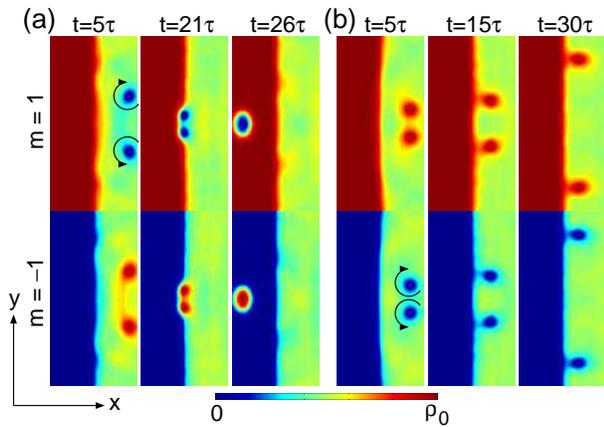}
\caption{
(Color online) Time evolution of a system in which a half-quantum vortex
dipole is imprinted to the polar side of the initial state.
In (a), singly-quantized vortices are imprinted only to the $m = 1$ wave
function, and in (b), they are imprinted only to the $m = -1$ wave
function.
The distances between the vortices in the initial vortex dipole are (a)
$\Delta = 6.4\xi_s$ and (b) $\Delta = 3.2\xi_s$.
The arrows indicate the directions in which the vortices circulate.
The window of each panel is $-5 \xi_s < x < 5 \xi_s$ and $-10 \xi_s < y <
10 \xi_s$.
See the Supplemental Material for movies of the dynamics of $|\psi_1|^2$
in (a) and that of $|\psi_{-1}|^2$ in (b).
}
\label{f:hqv2}
\end{figure}
We examine the dynamics of a half-quantum vortex dipole that is created on
the polar side and moves toward the ferromagnetic side.
In Fig.~\ref{f:hqv2}(a), vortices are imprinted only to the $m = 1$
component, using Eq.~(\ref{phase}) with $z_{\pm} = 5\xi \mp 3.2 \xi_s$.
As the half-quantum vortex dipole moves in the $-x$ direction and
approaches the phase boundary, the distance between the vortex and
antivortex is reduced ($t = 21 \tau$ in Fig.~\ref{f:hqv2}(a)).
On the ferromagnetic side, the vortex and antivortex merge into a single
density hole occupied by the $m = -1$ component, which continues to move
in the $-x$ direction ($t = 26 \tau$ in Fig.~\ref{f:hqv2}(a)).
This is quite different from Fig.~\ref{f:p_to_f}(f), where the vortex and
antivortex survive on the ferromagnetic side.
The different behaviors originate from the different energy scales of the
vortices.
It follows from the core sizes that the energies of the singly quantized
vortices in Fig.~\ref{f:p_to_f}(a) are larger than those of the spin
vortices in Fig.~\ref{f:p_to_f}(f), whereas the energies of the
half-quantum vortices in Fig.~\ref{f:hqv2}(a) are smaller.
Thus, the half-quantum vortex dipole cannot transform into a spin vortex
dipole on the ferromagnetic side due to a lack of energy.

Figure~\ref{f:hqv2}(b) shows the case in which the phase is imprinted only
to the $m = -1$ initial state, using Eq.~(\ref{phase}) with $z_{\pm} =
5\xi \mp 1.6 \xi_s$.
In this case, the penetration of a vortex dipole is topologically
prohibited, since the vortices in the $m = -1$ component cannot be
smoothly connected to those in the $m = 1$ component.
The half-quantum vortex dipole therefore does not penetrate the phase
boundary, but instead the vortex and antivortex disintegrate and move
along the phase boundary in opposite directions, as shown in
Fig.~\ref{f:hqv2}(b).

\section{Conclusions}
\label{s:conc}

We have investigated the dynamics of vortex dipoles across the boundary
between the ferromagnetic and polar phases in a spin-1 BEC.
We numerically solved the Gross--Pitaevskii equation and found a rich
variety of spin dynamics.
When a singly-quantized vortex dipole is created on the ferromagnetic
side and propagated toward the polar side, it transforms into one of two
kinds of spin-vortex dipoles, depending on its velocity.
For a large velocity, ferromagnetic-core vortex dipoles are created in the
$m = 1$ and $-1$ components, which exhibit the leapfrogging dynamics, as
shown in Figs.~\ref{f:f_to_p1} and \ref{f:mag}.
For a small velocity, a half-quantum vortex dipole is generated, as shown
in Fig.~\ref{f:hqv}.
When a singly-quantized vortex dipole created on the polar side moves into
the ferromagnetic side, it transforms into a vortex dipole whose
ferromagnetic cores are surrounded by half-quantum vortices, as shown in
Figs.~\ref{f:p_to_f}(a)--\ref{f:p_to_f}(f).
This state also exhibits the leapfrogging behavior on a different
quantization axis, as shown in Fig.~\ref{f:p_to_f}(g).
When a half-quantum vortex dipole is created on the polar side and
propagated to the ferromagnetic side, it coalesces into a low energy
droplet or is turned away at the phase boundary, as shown in
Fig.~\ref{f:hqv2}.

We have considered an ideal system that is infinite and homogeneous.
In a realistic experimental system confined in a trapping potential, the
system size must be much larger than a vortex dipole and its trajectory.
An elongated oblate BEC, as used in the Berkeley experiment~\cite{Sadler},
would be suitable.
An initial vortex dipole can be created by an external laser beam shifting
in a BEC~\cite{Neely,AioiX}.
A half-quantum vortex-dipole, as shown in Fig.~\ref{f:hqv2}, may be
created by an $m$-dependent external potential~\cite{Ji,Chiba}.

Our study presents an example of the dynamical transformation of
topological defects, in which topological structures are dynamically
changed.
It will also be interesting to consider collisions of topological
defects.
During collisions, topological structures may be dynamically changed, and
different topological structures may be scattered after the collisions,
which might be relevant to the collisions of elementary particles.

\begin{acknowledgments}
This work was supported by JSPS KAKENHI Grant Number 26400414 and by
KAKENHI (No. 25103007, ``Fluctuation \& Structure'') from MEXT, Japan.
\end{acknowledgments}

\end{document}